\documentclass[sigconf]{acmart}

\usepackage{booktabs} 
\usepackage{graphicx}
\newcommand{\indep}{\raisebox{0.05em}{\rotatebox[origin=c]{90}{$\models$}}}

\usepackage{array}
\setcopyright{rightsretained}

\acmDOI{}

\acmISBN{}

\acmConference[SIGIR 2018 eCom]{ACM SIGIR Workshop on eCommerce}{July 2018}{Ann Arbor, Michigan, USA} 
\acmYear{2018}
\copyrightyear{2018}

\acmSubmissionID{}

\begin{document}
\title{End-to-End Neural Ranking for eCommerce Product Search}
\subtitle{An application of task models and textual embeddings}

\author{Eliot P. Brenner}
\authornote{Corresponding Author.}
\orcid{0000-0002-9176-0595}
\affiliation{%
  \institution{Jet.com/Walmart Labs}
  \streetaddress{221 River Street}
  \city{Hoboken}
  \state{NJ}
  \postcode{07030}
}
\email{eliot.brenner@jet.com}

\author{Jun (Raymond) Zhao}
  \affiliation{%
  	\institution{Jet.com/Walmart Labs}
  	\streetaddress{221 River Street}
  	\city{Hoboken}
  	\state{NJ}
  	\postcode{07030}
}
\email{raymond@jet.com}

\author{Aliasgar Kutiyanawala}
 \affiliation{%
	\institution{Jet.com/Walmart Labs}
	\streetaddress{221 River Street}
	\city{Hoboken}
	\state{NJ}
	\postcode{07030}
}
\email{aliasgar@jet.com}

\author{Zheng (John) Yan}
\affiliation{%
	\institution{Jet.com/Walmart Labs}
	\streetaddress{221 River Street}
	\city{Hoboken}
	\state{NJ}
	\postcode{07030}
}
\email{john@jet.com}

\renewcommand{\shortauthors}{E. Brenner, J. Zhao, A. Kutiyanawala}

\begin{abstract}
We consider the problem of retrieving and ranking items in an eCommerce catalog, often called \textit{SKU}s, in order of relevance to a user-issued query.  The input data for the ranking are the texts of the queries and textual fields of the SKUs indexed in the catalog.  We review the ways in which this problem both resembles and differs from the problems of information retrieval (IR) in the context of web search,
which is the context typically assumed in the IR literature.  The differences between the product-search problem and the IR problem of web search necessitate a different approach in terms of both models and datasets.  We first review the recent state-of-the-art models for web search IR, focusing on
the CLSM of \cite{shen2014latent} as a representative of one type, which we call the \textit{distributed} type, and the kernel pooling model of \cite{xiong2017end}, as a representative of another type, which we call the \textit{local-interaction} type.  The different types of relevance models developed for IR have complementary advantages and disadvantages when applied to eCommerce product search.  Further, we explain why the conventional methods for dataset construction employed in the IR literature fail to produce data which suffices for training or evaluation of models for eCommerce product search.  We explain how our own approach,
applying \textit{task modeling} techniques to the click-through logs of an eCommerce site, enables the construction of a large-scale dataset for training and robust benchmarking of relevance models.  Our experiments consist of applying several of the models from the IR literature to our own dataset.  Empirically, we have established that, when applied to our dataset, certain models of \textit{local-interaction} type reduce ranking errors by one-third compared to the baseline system (tf---idf).  Applied to our dataset, the \textit{distributed} models fail to outperform the baseline.  As a basis for a deployed system, the \textit{distributed} models have several
advantages, computationally, over the \textit{local-interaction} models.  This motivates an ongoing program of work, which we outline at the conclusion of the paper.
\end{abstract}



\maketitle

\section{Introduction}
Currently deployed systems for eCommerce product search tend to use
inverted-index based retrieval, as implemented in Elasticsearch \cite{gormley2015elasticsearch}
or Solr \cite{smiley2015apache}.  For ranking, these systems
typically use \textit{legacy} relevance functions such as tf---idf \cite{WPtf-idf} or
Okpai BM25 \cite{robertson2009probabilistic}, as implemented in these search systems.
Such relevance functions are based on exact ("hard") matches of tokens, rather than semantic ("soft") matches, are insensitive to word order, and have hard-coded, rather than learned weights.
On the one hand, their simplicity makes legacy relevance functions scalable and easy-to-implement.  One the other hand, they are found to be inadequate in practice for fine-grained ranking of search results.  Typically, in order to achieve rankings of search results that are acceptable for presentation to the
user, eCommerce sites overlay on top of the legacy relevance function score, a variety of handcrafted filters (using structured data fields) as well
as hard-coded rules for specific queries.  In some cases, eCommerce sites
are able to develop intricate and specialized proprietary NLP systems, referred
to as Query-SKU Understanding (QSU) Systems, for analyzing and matching relevant SKUs to queries.  QSU
systems, while potentially very effective at addressing the shortcomings of legacy relevance scores, require a 
some degree of domain-specific knowledge to engineer \cite{kutiyanawala2018ontology}.  Because of concept drift, the maintenance of QSU systems demands a long-term commitment of analyst and programmer labor.  
As a result of these scalability issues, QSU systems are within reach only for the very largest eCommerce companies with abundant resources.

Recently, the field of neural IR (NIR) , has shown great promise to overturn this
state of affairs.   The approach of NIR to ranking
differs from the aforementioned QSU systems in that it
\textit{learns vector-space representations} of both
queries and SKUs which facilitate learning-to-rank (LTR) models to address
the task of relevance ranking in an end-to-end manner.  NIR,
if successfully applied to eCommerce, can allow any company with access to commodity GPUs and abunant
user click-through logs to build an accurate and robust model for ranking search
results at lower cost over the long-term compared to a QSU system. For a current and comprehensive review
of the field of NIR, see \cite{mitra2017neural}.  

Our aim in this paper is to provide further theoretical justification and empirical
evidence that fresh ideas and techniques are needed to make Neural IR a practical
alternative to legacy relevance and rule-based systems for eCommerce search.  Based
on the results of model training which we present, we delineate
a handful of ideas which appear promising so far and deserve further development.

\section{Relation to Previous Work in Neural IR}
\label{sec:neural_ir}
The field of NIR shares
its origin with the more general field of neural NLP \cite{goldberg2017nnnlp} in the \textit{word embeddings} work of word2vec \cite{le2014distributed} and its variants.  From there, though,
the field diverges from the more general stream of neural NLP in a variety
of ways.  The task of NIR, as compared with other
widely known branches such as topic modeling, document clustering,
machine translation and automatic summarization, consists
of \textit{matching} texts from one collection (the queries) with texts from another collection (webpages or SKUs, or other
corpus entries, as the case may be).  The specific challenges and innovations of NIR in terms of models are driven by the fact that entries from the two collections (\textit{queries} versus \textit{documents}) are of a very different nature from one another, both in length, and internal structure
and semantics.  

In terms of datasets, the notion
of \textit{relevance} has to be defined in relation
to a specific IR task and the population which will use the system.  In contrast to the way many general NLP models can be trained and evaluated on publicly available labeled benchmark corpora, a NIR model has to be trained on a datasest tailored to reflect the information needs of of the population the system is meant to serve.  In order to produce such task-specific datasets in a reliable and scalable manner, practitioners need to go beyond traditional methods such as expert labeling and simple statistical aggregations.  The solution has been to take "crowdsourcing" to its logical conclusion and to use the "big data" of user logs to extract relevance judgments.  This has led NIR to depend for scalability on
 the field of Click Models, which are discussed at great length in 
 the context of web search in the book \cite{chuklin2015click}.
 
While NIR has great promise for application to eCommerce, the existing
NIR literature has thus far been biased towards the web search problem,
relegating eCommerce product search to "niche" field status.
We hope to remedy this situation by demonstrating the need
for radical innovation in the field of NIR to address the challenges
of eCommerce product search.  

One of the most important differences is that
for product search, as compared to web search, both the intent and vocabulary of queries tend
to be more restricted and "predictable".  
For example, when typing into a commercial web search
engine, people can be expected to search for \textit{any} type of information.
Generally speaking, customers on a particular eCommerce site
are only looking to satisfy only one type of information need, namely to retrieve
a list of SKUs in the catalog meeting certain criteria.  
Users, both customers and merchants, are highly motivated by economic
concerns (time and money spent) to craft queries and SKU text fields to facilitate the surfacing of relevant content,
for example, by training themselves to pack popular and descriptive keywords into queries and SKUs.
As a consequence, the non-neural baselines, such as tf---idf, tend 
to achieve higher scores on IR metrics in product search datasets than in web search.  An NIR model, when applied to product search as opposed to web search, has a higher bar to clear
to justify the added system complexity.  

Further, the lift NIR provides over tf---idf is largely in detecting semantic, non-exact matches.  For example, consider a user
searching the web with the query "king's castle".  This user will likely have her
information need met with a document about the "monarch's castle" or even possibly "queen's castle",
even if it does not explicitly mention "king".  In contrast, consider
the user issuing the query "king bed" on an eCommerce site.   She would likely
consider a "Monarch bed" \footnote{For the sake of this example, we are assuming
that "Monarch" is a brand of bed, only some of which are "king" (size) beds.} irrelevant
unless that bed is also "king", and a "queen bed" even more irrelevant.  An approach based on word2vec or Glove \cite{pennington2014glove} vectors would likely
consider the words "king", "queen" and "monarch" all similar to one another based on the distributional hypothesis.  The baseline systems deployed on eCommerce websites often deal with semantic similarity by
incorporating analyzers with handcrafted synonym lists built in, an 
approach which is completely infeasible for open-domain web search.  This further
blunts the positive impact of learned semantic similarity relationships.  

Another difference between "general" IR for web-search and IR specialized
for eCommerce product search is that in the latter the relevance landscape
is simutaneously "flatter" and more "jagged".  With regard to "flatness", consider a very popular web search
for 2017 (according to \cite{Google2018}):
"how to make solar eclipse glasses".  Although
the query expresses a very specific search intent, it is likely
that the user making it has other, related information needs.  
Consequently, a good search engine results page (SERP), could be composed entirely of links to instructions on making solar eclipse glasses, which while completely
relevant, could be redundant.  A better SERP would be composed of a mixture of the former, and of links
to retailers selling materials for making such glasses,
to instructions on how to use the glasses, and to the best locations for viewing the upcoming
eclipse, all which are likely relevant to some degree to the user's infromation need and improve SERP diversity.  
In contrast, consider the situation for a more typical eCommerce query: "tv remote".
A good product list view (PLV) page for the search "tv remote" would display \textit{exclusively} SKUs which are tv remotes, and \textit{no} SKUs which are tvs.  Further, all the "tv remote" SKUs are equally \textit{relevant}
to the query, though some might be more popular and engaging than others.  With regard
to "jaggedness", consider the query "desk chair": 
any "desk \textit{with} chair" SKUs would be considered completely irrelevant and do
not belong anywhere on the PLV page, spite of the very small lexical
difference between the queries "desk chair" and "desk with chair".  In web search,
by contrast, documents which are relevant for a given query tend
to remain partially relevant for queries which are lexically similar to the original query.  If
the NIR system is to compete with currently deployed rule-based systems, it is of great importance
for a dataset for training and benchmarking NIR models for product search
to incorporate such "adversarial" examples prevalently. 

Complications arise when we estimate relevance from the click-through logs of an eCommerce
site, as compared to web search logs.  Price, image quality
are factors comparable in importance to relevance in driving customer click
behavior. As an illustration, consider the situation
in which the top-ranked result on a SERP or PLV page
for a query has a lower click-through rate (CTR) than
the documents or SKUs at lower ranks.  
In the web SERP case, the low CTR sends a strong signal that the document is irrelevant to the query, but in
the eCommerce PLV it may have other other explanations, for example, the SKU's being mispriced
or having an inferior brand or image.  We will address this point in much more detail in Section \ref{sec:click_models} below.

Another factor which assumes more importance in the field of eCommerce search
versus web search is the difference between using the NIR model
for \textit{ranking only} and using it for \textit{retrieval \textbf{and} ranking}.  In the former
scenario, the model is applied at runtime as the last step of a pipeline or "cascade", where
earlier steps of the pipeline use cruder but computationally faster techniques
(such as inverted indexes and tf---idf) to identify a small subcorpus (of say a few hundred SKUs)
as potential entries on the PLV page, and the NIR model only re-ranks
this subcorpus (see e.g. \cite{tu2017exploration}).  In the latter scenario, the model, or more precisely,
approximate nearest neighbor (ANN) methods acting on vector
representations derived from the model, both select and rank the search
results for the PLV page from the entire corpus.  The latter scenario, retrieval
and ranking as one step, is more desirable but (see \S\S\ref{sec:NIR_models} and \ref{sec:experimental_results} below) poses additional challenges
for both algorithm developers and engineers.  The possibility of using the NIR model for retrieval, as opposed to mere re-ranking, receives relatively little attention in the web search literature, presumably because of its infeasibility: the corpus of web documents is so vast, in the trillions \cite{VentureBeat2013}, compared to only a few million items in each category of eCommerce SKUs.

\section{Neural Information Retrieval Models}
\label{sec:NIR_models}
In terms of the high-level classification of Machine Learning tasks, which includes such categories as
"classification" and "regression", Neural Information Retrieval (NIR) falls
under the  Learning to Rank (LTR) category (see \cite{li2014learning} for a comprehensive survey).  An LTR algorithm uses an objective function defined over possible "queries" $q\in Q$ and lists of "documents" $d\in D$, to learn a model that, when applied to an new previously unseen query, uses
the features of the both the query and documents to "optimally" order the documents for
that query.  For a more formal definition, see \S1.3 of \cite{li2014learning}.
As explained in \S2.3.2 of \cite{li2014learning}, it is common practice
in IR to approximate the general LTR task with a binary classification
task.  For the sake of simplicity, we follow this so-called "pairwise approach" to LTR.  Namely, we first
consider LTR algorithms whose orderings are induced from a scoring function $f: Q\times D\rightarrow \mathbf{R}$ in the sense that the ordering for $q$ consists of sorting
$D$ in the order $d_1,\ldots d_N$ satisfying $f(q,d_1)\geq f(q,d_2)\geq \cdots f(q,d_N)$.  Next,
we train and evaluate the LTR model by presenting it with triples
$(q,d_{\rm rel}, d_{\rm irrel})\in Q\times D \times D$, where $d_{\rm rel}$ is deemed to be more relevant
to $q$ than $d_{\rm irrel}$: the binary classification task amounts to assigning scores $f$
so that $f(q,d_{\rm rel}) > f(q,d_{\rm irrel})$.

There are many ways of classifying the popular types of NIR models, but
the way that we find most fundamental and useful for our purposes is the classification
into \textit{distributed} and \textit{local-interaction} models.
The distinction between the two types of models lies in the following architectural difference.
The first type of model first transforms $d$ and $q$ into "distributed" vector
representations $v_D(d)\in V_D$ and $v_Q(q)\in V_Q$ of fixed dimension, and only after that feeds the representations into an LTR model.  The second type of model
never forms a fixed-dimensional "distributed" vector representation of document or
query, and instead forms a "interaction" matrix representation $\langle i(q), i(d) \rangle$ of the pair $(q,d)$.  The matrix $\langle i(q), i(d) \rangle$ 
is sensitive only to local, not global, interactions between $q$ and $d$,
and the model subsequently uses $\langle i(q), i(d) \rangle$ as the input
to the LTR model.  More formally, in the notation of Table \ref{tab:notation},
the score function $f$ takes the following form in the \textit{distributed} case:
\begin{equation}\label{eqn:distributed_nir_score}
f(q,d) = r\left( v_Q(q), v_D(d) \right) = r\left( g_Q(i(q)), g_D(i(d)) \right),
\end{equation}
and the following form in the \textit{local-interaction} case:
\begin{equation}
\label{eqn:local_interaction_nir_score}
f(q,d) = r\big( \langle i(q), i(d)\rangle  \big).
\end{equation}
Thus, the interaction between $q$ and $d$ occurs at a \textit{global} level in the distributed
case and at the \textit{local} level in the local-interaction case.
The NIR models are distinguished individually from others of the same type by the choices
of building blocks in Table \ref{tab:notation}.\footnote{Although the dimension of word2vec is a hyperparameter,	we make the conventional choice that word-vectors have
	dimension $300$ in the examples to keep the number of symbols to a minimum.}  
In Tables \ref{tab:distributed_nir_models} and \ref{tab:interaction_nir_models}, we show how to obtain
some common NIR models in the literature by making specific choices of the building blocks.
Note that since $w$ determines $i$, and $g,i$ together determine $v$, to specify a local-interaction model
(Table \ref{tab:interaction_nir_models}) we only need to specify the mappings $w,r$, and to specify a distributed NIR model (Table \ref{tab:distributed_nir_models}), we need only additionally specify $g$. 
We remark that the distributed word embeddings in the "Siamese" and Kernel Pooling models
are implicitly assumed to be normalized to have $2$-norm one, which allows
the application the dot product $\langle\cdot, \cdot \rangle$ to compute the cosine similarity.

The classification of NIR models in Tables \ref{tab:distributed_nir_models} and \ref{tab:interaction_nir_models}
is not meant to be an exhaustive list of all models found in the literature: for a more
comprehensive exposition, see for example the baselines section \S4.3 of \cite{xiong2017end}.
Laying out the models in this fashion facilitates a determination of which
parts of the "space" of possible models remain unexplored.  For example, we
see the possibility of forming a new "hybrid" by using the distributed word embeddings layer $w$ from the Kernel Pooling model \cite{xiong2017end}, followed by the representation layer $g$ from CLSM \cite{shen2014latent},
and we have called this distributed model, apparently nowhere considered
in the existing literature "Siamese".  Substituting into \eqref{eqn:distributed_nir_score},
we obtain the following formula for the score $f$ of the Siamese model:
\begin{equation}\label{eqn:siamese}
f(q,d) = \langle \mathrm{mlp}_1\circ \mathrm{cnn}_1(i(q)),
 \mathrm{mlp}_1\circ \mathrm{cnn}_1(i(d))\rangle.
\end{equation}
As an example missing from the space of possible local-interaction
models, we have not seen any consideration of the "hybrid" architecture obtained
by using the local-interaction layer from the Kernel-Pooling model and
a position-aware LTR layer such as a convolution neural net.  Substituting
into \eqref{eqn:local_interaction_nir_score}, the score function would be
\begin{equation}\label{eqn:interaction_general_ltr}
f(q,d) = \mathrm{mlp}\circ \mathrm{cnn}\big( 
 \langle i(q), i(d)\rangle  
 \big),
\end{equation}
where in both \eqref{eqn:siamese} and \eqref{eqn:interaction_general_ltr},
$i$ is the embedding layer obtained induced by a trainable word embedding $w$. 
These are just a couple of the other new architectures which
could be formed in this way: we just highlight these two because they 
seem especially promising.

The main benefit of distinguishing between distributed and local-interaction
models is that a distributed architecture enables retrieval-and-ranking, whereas a local-interaction architecture
restricts the model to re-ranking the results of a separate retrieval mechanism.  See
the last paragraph of Section \ref{sec:neural_ir} for an explanation of this distinction.
From a computational complexity perspective, the reason for this is that
the task of precomputing and storing all the representations involved in computing
the score is, for a distributed model, $O(|Q|+|D|)$ in space and time, but for a local-interaction model, $O(|Q||D|)$.  
(Note that we need to compute the representations \textit{which are the inputs
to} $r$, namely, in the distributed case, the vectors $v_Q(q)$, $v_D(d)$,
and in the local-interaction case, the matrices $\langle i(q),i(d)\rangle$: computing the variable-length representations
$i(d)$ and $i(q)$ alone will not suffice in either case).  
Further, in the distributed model case, assuming the LTR function $r(\cdot)$ is chosen to be the dot-product $\langle\cdot,\cdot\rangle$,
the retrieval step can be implemented by ANN.  This is the reason
for the choice of $r$ as $\langle\cdot, \cdot\rangle$
in defining the "Siamese" model.  For practical implementations of 
ANN using Elasticsearch at industrial scale, see \cite{rygl2017semantic} for
a general text-data use-case, and \cite{mu2018visualsearch} for an image-data use-case
in an e-Commerce context.  We will return to this point in \S\ref{sec:conclusions}.
\begin{table*}
	\caption{Notation for Building Blocks of NIR models}
	\label{tab:notation}
	\begin{tabular}{cll}
		\toprule
		Symbol & Meaning & Examples or Formulas\\
		\midrule
		$\langle\cdot,\cdot \rangle$ & dot-product $\langle A,B\rangle = A\cdot B^T$ for vectors or matrices $A,B$ & $A\in \mathbf{R}^{m\times k}, B\in\mathbf{R}^{n\times k}\Rightarrow \langle A,B\rangle \in\mathbf{R}^{m\times n}$, $\langle A,B \rangle_{i,j} = \sum_{\ell=1}^k A_{i,\ell}B_{j,\ell}$ \\
		$[\ldots]$ & Matrix Augmentation (concatenation) & $A_1,\ldots, A_K\in \mathbf{R}^{M\times N}\Rightarrow [A_1,\ldots, A_k]\in \mathbf{R}^{M\times KN}$\\
		$\mathrm{mlp}_k$ & fully-connected ($k$-layer) perceptron & $\mathrm{mlp}_2(x)=\tanh(W_2\cdot\tanh(W_1\cdot x+b_1)+b_2)$, $W_i$ weight matrices \\
		$\mathrm{cnn}_k$ & $k$-layer convolutional nn (with possible max-pooling) & $\mathrm{cnn}_1(x) = \tanh(W_c\cdot x)$, $W_c$ convolutional weight matrix\\[.3cm]
		$w$ & word or embedding &  \begin{tabular}{@{}l@{}}word hashing of \cite{huang2013learning} \\ 	mapping derived from word2vec, e.g. token $t\mapsto w(t)\in \mathbf{R}^{300}$\end{tabular}
	 \\[.4cm]
		$i$ & local document embedding & $i: [t_1,\ldots, t_k]\mapsto [\underbrace{w(t_1)^T,\ldots,w(t_k')^T}_{k'=\min(k,N)},
		\underbrace{\mathbf{0}^T,\ldots, \mathbf{0}^T}_{N-k'\;\mathrm{times}}]\in \mathbf{R}^{300\times N}$ \\[0.6cm]
		$N_{\{Q|D\}}$ & Dimension of local document embedding along token axis & $N_D=1000$, $N_Q=10$ for Duet model.\\
		$V_{\{Q|D\}}$ & Vector space for \textbf{distributed} representations of $D$ or $Q$ & $V_Q=\mathbf{R}^{300}$, $V_D=\mathbf{R}^{300\times 899}$ for Duet Model\\
		$V$ & Vector space for \textbf{distributed} representations of $D$ and $Q$ & as above,
		when $V_Q=V_D$.\\ 
		 $g_{\{Q|D\}}$   & parameterized mapping of $i(\{q|d\})$ to $v_{\{q|d\}}(\{q|d\})\in V_{\{Q|D\}}$ & $\mathrm{mlp}$; $\mathrm{cnn}$; $\mathrm{mlp}\circ\mathrm{cnn}$; $\mathrm{mlp}\circ\sigma_1$\\
		 $g$   & parameterized mapping of $i(d)$, $i(q)$ to $v \in V$ & special
		 case of above, when $V_Q=V_D$, and 
		 $g_Q=g_D$\\
		    $v_{\{Q|D\}}$  & $g_{\{q|d\}}\circ i$, global document embedding of $q$ or $d$ into $V_{\{Q|D\}}$ & composition or previous mappings\\
		     $v$  & $g\circ i$, global document embedding of $q$ or $d$ into $V$ &  special case when $V_Q=V_D$, and $g_Q=g_D$, $v_Q=v_D$\\
		       $r$ & LTR model mapping $\langle i(q),i(d)\rangle$ or $(v_Q(q), v_D(d))\mapsto\mathbf{R}$   & $\langle\cdot,\cdot\rangle$ \cite{shen2014latent}; 
		       $\mathrm{mlp}_3\circ \bigodot$ \cite{mitra2017learning}; $\mathrm{mlp}_1\circ\mathbf{K}$ \cite{xiong2017end} \\
		       $\sigma_1$ & summation along $1$-axis, mapping $\mathbf{R}^{m\times n} \rightarrow \mathbf{R}^m$ & $\sigma_1: \mathbf{R}^{m \times n}\rightarrow \mathbf{R}^m$, $\sigma_1(A)_i=\sum_{j=1}^n A_{i,j}$ \\ 
		       $\bigodot$ & Hadamard (elementwise) product (of matrices, tensors)  & $A,B\in \mathbf{R}^{m\times n}\Rightarrow A\bigodot B\in \mathrm{R}^{m\times n}$\\
		       $\bigodot$ & Broadcasting, followed by Hadamard (operator overloading) & $A \in R^{m}, B\in R^{m\times n}\Rightarrow A\bigodot B=[\underbrace{A,\ldots, A}_{n\;\textbf{times}}]\bigodot B$ \\
		       $\mathbf{K}$ & Kernel-pooling map of \cite{xiong2017end} & $\mathbf{K}(M_i) = [K_1,\ldots, K_K]$, $K_k$ RBF kernels with mean $\mu_k$. \\
		       $\phi$ & Kernel-pooling composed with soft TF map of \cite{xiong2017end} & $M\in\mathbf{R}^{N_Q\times N_D}\Rightarrow \phi(M)=\sum_{i=1}^{N_Q}\log \mathbf{K}(M_i)$\\
		\bottomrule
	\end{tabular}
\end{table*}

\begin{table*}
	\caption{Comparison of distributed NIR models in literature}
	\label{tab:distributed_nir_models}
	\begin{tabular}{clllllll}
		\toprule
		Model & $w$ &$\mathrm{dim}(w)$& $g_Q$ &$g_D$& $\mathrm{dim}\left(V_Q\right)$ & $\mathrm{dim}\big( V_D \big)$ & $r$ \\
				\midrule
		DSSM \cite{huang2013learning} & "word hashing", see \cite{huang2013learning} & $30,000$ & $\mathrm{mlp}_3\circ \sigma_1$ & $=g_Q$&$128$ & $128$ & $\langle\cdot,\cdot\rangle$ \\
		CLSM \cite{shen2014latent} & "word hashing", see \cite{huang2013learning} & $30,000$ & $\mathrm{mlp}_1\circ \mathrm{cnn}_1$& $=g_Q$ &$128$ & $128$ & $\langle\cdot,\cdot\rangle$ \\
		Duet \cite{mitra2017learning} (distributed side) & "word hashing", see \cite{huang2013learning} & $2000$ & $\mathrm{mlp}_1\circ\mathrm{cnn}_1$& $\mathrm{cnn}_2$& $300$& $300\times 899$ & $\mathrm{mlp}_3\circ\bigodot$ \\
		"Siamese" (ours) & distributed word embeddings & $300$ & $\mathrm{mlp}_1 \circ \mathrm{cnn}_1$ & $=g_Q$ & $32-512$ & $32-512$ & $\langle\cdot,\cdot\rangle$\\ 
		\bottomrule
	\end{tabular}
\end{table*}
\begin{table*}
	\caption{Comparison of local-interaction NIR models in literature}
	\label{tab:interaction_nir_models}
	\begin{tabular}{clll}
   \toprule
   	Model & $w$ & $\mathrm{dim}(w)$ & $r$\\
   			\midrule
   tf---idf & one-hot encoding, weighted by $\sqrt{\mathrm{idf}}$ & $|\mathrm{Vocabulary}|$ & $\sigma_1:\,\mathbf{R}^{|\mathrm{Vocabulary}|}_+\rightarrow \mathbf{R}_+$ \\
   Duet \cite{mitra2017learning} (local side) & one-hot encoding & $|\mathrm{Vocabulary}|$ &  $\mathrm{mlp}_3\circ\mathrm{cnn}_1$\\
   Kernel-pooling model \cite{xiong2017end} & distributed word embeddings & $300$ & $\mathrm{mlp}_1\circ \phi$\\
   \bottomrule	
\end{tabular}
\end{table*}

\section{From Click Models to Task Models}
\label{sec:click_models}
\subsection{Click Models}
\label{subsec:click_models}
The purpose of click models is to extract, from observed variables, an estimate of latent variables.
The observed variables generally include the sequence
of queries, PLVs/SERPs and clicks, and may also include
hovers, add-to-carts (ATCs), and other browser interactions recorded in the
site's web-logs.  The main latent variables are the relevance
and attractiveness of an item to a user.  A click model historically takes the form of a probabilistic
graphical model called a Bayesian Network (BN), whose structure
is represented by a Directed Acyclic Graph (DAG), though more recent works have introduced
other types of probabilistic model including recurrent \cite{borisov2016neural} and adversarial \cite{moore2018modeling}
neural networks.  The click model embodies certain
assumptions about how users behave on the site.   
We are going to focus on the commonalities of the various click models
rather than their individual differences, because
our aim in discussing them is to motivate
another type of model called \textit{task models},
which will be used to construct the experimental dataset in Section \ref{sec:dataset_construction}.

\begin{figure}
\includegraphics[height=1.4in, width=1.65in]{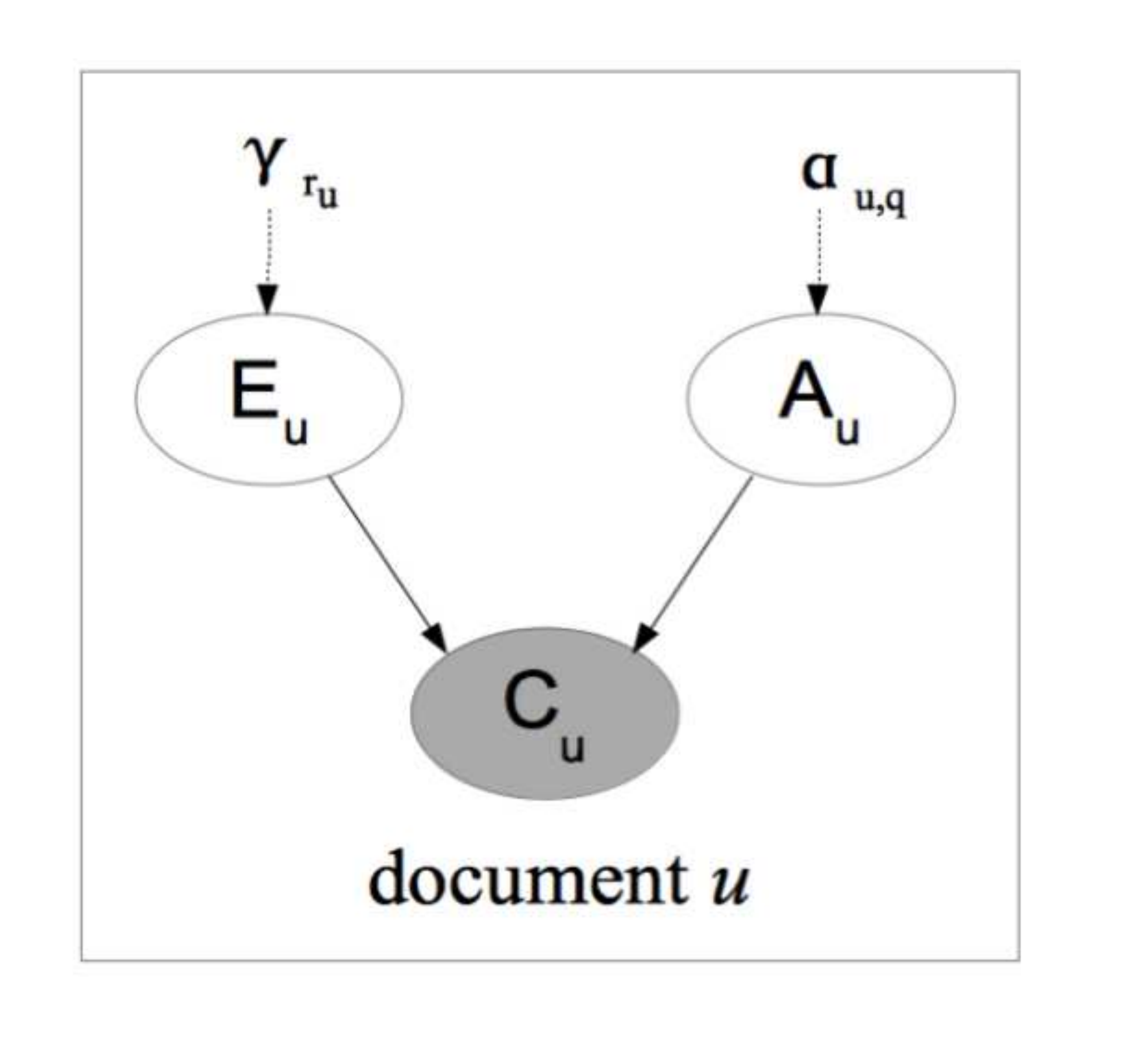}
\caption{Template-based representation of the PBM click model. \label{fig:pbm_dag}}
\end{figure}

To model the stochasticity inherent in the user browsing process, 
click models adopt the machinery of probability theory and conceptualize the click event,
 as well as related events such as examinations, ATCs, etc.,
 as the outcome of a (Bernoulli) random variable.  The three fundamental
 events for describing user interaction with a SKU $u$ on a PLV are denoted as follows: 
 \begin{enumerate}
 	\item  The user \textit{examining} the SKU, denoted by $E_u$;
 	\item The user being sufficiently \textit{attracted} by the SKU "tile" to click it, denoted by $A_u$;
 	\item  The user \textit{clicking} on the SKU, by $C_u$.
 \end{enumerate}
The most basic assumption relating these three events is the following 
\textbf{Examination Hypothesis} (Equation (3.4), p. 10 of \cite{chuklin2015click}):
\begin{equation}\label{eqn:examination_hypothesis}
	 \text{\textbf{EH}}\quad C_u=1\Leftrightarrow E_u=1\;\text{and}\; A_u=1.
	\end{equation}
The \textbf{EH} is universally shared by click models, in view of the observation that any click which the user makes \textit{without} 
examining the SKU is just "noise" because it
cannot convey any information about the SKU's relevance.
In addition, most of the standard click models, including the Position Based (PBM) and Cascade (CM) Models,
also incorporate the following \textbf{Independence Hypothesis}:
\begin{equation}\label{eqn:independence_hypothesis}
\textbf{\textbf{IH}}\quad E_u\, \indep \,A_u.
\end{equation}    
The \textbf{IH} appears reasonable because the attractiveness of a SKU
represents an inherent property of the query-SKU pair, whereas
whether the SKU is examined is an event contingent on a particular presentation on the PLV and the individual
user's behavior.  This means that $E_u$ and $A_u$ should not be able to influence one another,
as the \textbf{IH} claims.  Adding the following two ingredients to the \textbf{EH} \eqref{eqn:examination_hypothesis}
and the \textbf{IH} \eqref{eqn:independence_hypothesis}, we obtain a functional, if minimal,
click model
\begin{enumerate}
	\item A probabilistic description of how the user interacts
with the PLV: i.e., at a minimum, a probabilistic model of the variables $E_u$.
\item A  parameterization of the distributions underlying the variables  $A_u$. 
\end{enumerate}
 We can specify a simple click model by carrying out (1)--(2) as follows:
 \begin{enumerate}
\item $P(E_u=1)$ depends only on the rank of $u$, and is given by a single parameter $\gamma_r$
for each rank, so that $P(E_{u_r})=:\gamma_r\in[0,1]$.
\item $P(A_u=1|E_u=1)=:\alpha_{u,q}\in[0,1]$, where there
is an independent Bernoulli parameter $\alpha_{u,q}$ for each pair of query and SKU.  
\end{enumerate}
The click model obtained by specifying  (1)--(2) as above is called the PBM.  For a template-based representation of the PBM, see
 Figure \ref{fig:pbm_dag}.  For further detail on the PBM
and other BN click models, see \cite{chuklin2015click}, and for
interpretation of template-based representations of BNs see Chapter 6 of \cite{koller2009probabilistic}.

The parameter estimation of click model BN's relies on the following
consequence of the \textbf{EH} \eqref{eqn:examination_hypothesis}:
\begin{equation}\label{eqn:likelihood_implication}
\left\{E_u=E_{u'}=1\;\& \;\, C_u=0\;\&\; C_{u'}=1\right\} \rightsquigarrow \alpha_{u',q} > \alpha_{u,q},
\end{equation}
where in \eqref{eqn:likelihood_implication}, the symbol $\rightsquigarrow$
means "increases the likelihood that".  This still leaves
open the question of \textit{how to connect attractiveness to relevance},
which click models need to do in order to fulfill their purpose,
namely extracting relevance judgments from the click logs.
The simplest way of making the connection is to assume that an item (web page or SKU) is attractive
if and only if it is relevant, but this assumption is obviously implausible even
in the web search case, and all but the simplest click models reject it.
A more sophisticated approach, which is the option taken by click models such as  DCM, CCD, DBN,
is to define another Bernoulli random variable
\begin{enumerate}
\setcounter{enumi}{3}
\item $S_u$, the user having her information need satisfied by SKU $u$,
\end{enumerate}
satisfying the following \textbf{Click Hypothesis}: 
\begin{equation}\label{eqn:click_satisfaction_assumption}
\textbf{CH}\;C_u=0 \Rightarrow S_u=0.
\end{equation}
The \textbf{CH} is formally analogous to the 
 \textbf{EH}, \eqref{eqn:examination_hypothesis}, since just as the \textbf{EH}
says that only $u$ for which $E_u=1$ are eligible to have $C_u=1$, regardless of their
inherent attractiveness $\alpha_{u,q}$, the \textbf{CH} says that only $u$ for which $C_u=1$ are eligible to have 
$S_u=1$, regardless of their inherent relevance $\sigma_{u,q}$.  The justification
of the \textbf{CH} is that the user
cannot know if the document is truly relevant, and thus
cannot have her information need satisfied by the document, unless she clicks on the document's link displayed
on the SERP.  According to the \textbf{CH}, in order to completely describe $S_u$,
we need only specify its conditional distribution when $C_u=1$.  We will follow the above click models
by specifying (in continuation of (1)-(2) above):
\begin{enumerate}
	\setcounter{enumi}{2}
	\item $P(S_u=1|C_u=1)=:\sigma_{u,q}\in[0,1]$, where there is an independent Bernoulli parameter $\sigma_{u,q}$
	for each pair of query and SKU.
\end{enumerate} 
Since $S_u$ is latent, to allow estimation of the paramters $\sigma_{u,q}$,
we need to connect $S_u$ to the observed variables $C_u$, and the connection is
generally made through the examination variables $\{E_u'\}$, by an assumption
such as the following:
\[
P(E_{u'}=1|S_u=1)\neq P(E_{u'}=1|S_u=0),
\]
where $u'\neq u$ is a SKU encountered \textit{after} $u$ in the process of browsing
the SERP/PLV. 

\textit{Our reason for highlighting the \textbf{CH} is that we have found
that the \textbf{CH} limits the applicability of these models in practice.}
Consider the following \textbf{Implication Assumption}:
\begin{equation}\label{eqn:implication_assumption}
\textbf{IA}\quad \sigma_{u,q}\; \text{is high} \Rightarrow \alpha_{u,q}\;\text{is high}.
\end{equation}
Clearly the \textbf{IA} is not \textit{always} true, even in the SERP case, because,
for example, the search engine may have done a poor job of producing the snippet
for a relevant document.  But in order for
parameter tuning to converge in a data-efficient manner, the \textbf{AI} must be 
\textit{predominantly} true.  The reason
for this is that, according to the \textbf{CH} \eqref{eqn:click_satisfaction_assumption},
the variable $C_u$ acts as a "censor" deleting random samples
for the variable $S_u$. For a fixed number $N$ of observations of a SKU
with fixed $\sigma_{u,q}$, the effective sample size for the empirical estimate $\hat{\sigma}_{u,q}$ 
is approximately $N\cdot\alpha_{u,q}$, so that as 
$\alpha_{u,q}\rightarrow 0_+$, the variance of $\hat{\sigma}_{u,q}$ is scaled up
by a factor of $1/\alpha_{u,q}\rightarrow\infty$ .  See
Chapter 19 of \cite{koller2009probabilistic} for a more comprehensive
discussion of \textit{values missing at random} and associated
phenomena.  We have already discussed in \S\ref{sec:neural_ir} that relevance
of a SKU is a \textit{necessary but not sufficient condition} for a high CTR,
and consequently the \textbf{IA}, \eqref{eqn:implication_assumption} is
frequently violated in practice for SKUs.
Empirically, we have observed poor performance of most click models (other than the PBM and UBM)
on eCommerce site click-through logs, and we attribute this to the failure of the \textbf{IA}.  Thus a suitable
modification of the usual click model framework is needed to extract relevance
judgments from click data.  This is the subject of \S\ref{sec:task_models} below.

\subsection{Relevance versus Attractiveness}
\label{sec:relevance}
At this point, we take a step back from our development
of the task model to address a question that the reader may already be asking:
given the complications of extracting relevance from click logs in the eCommerce
setting, why not completely forgo modeling \textit{relevance} and instead
model the \textit{attractiveness} of SKUs directly?  After all, it would seem
that the goal of search ranking in eCommerce is to present
the most engaging PLV which will result in the most clicks, ATCs, and purchases.
If a model can be trained to predict SKU attractiveness directly from query
and SKU features in an end-to-end manner,
that would seem to be sufficient and decrease the motivation to model relevance separately.

There are at least two practical reasons for wanting to model relevance separately from attractiveness.  The first is that relevance is the most stable among the factors
that affect attractiveness, namely price, customer taste, etc., all of which vary significantly over time.  
Armed with both a model of relevance and a separate
model of how relevance interacts with the more transient factors to impact attractiveness, the site manager 
can estimate the effects on attractiveness
that can be achieved by pulling various levers available to her, i.e., by modifying the transient factors (changing the price, or attempting to  alter customer taste through different marketing).  The second
is that the textual (word, query, and SKU-document) representations produced by modeling
relevance, for example, by applying any of the \textit{distributed}
models from \S\ref{sec:NIR_models}, have potential applications in related areas
such as recommendation, synonym identification, and automated ontology construction.
Allowing transient factors correlated with attractiveness, such
as the price and changing customer taste, to influence
these representations, would skew them in unpredictable and undesirable ways, limiting 
their utility.  We will return to the point of non-search applications of the representations in \S\ref{sec:conclusions}. 

\subsection{Task Models}
\label{subsec:task_models}
\begin{figure}
\includegraphics[height=1.8in, width=3.2in]{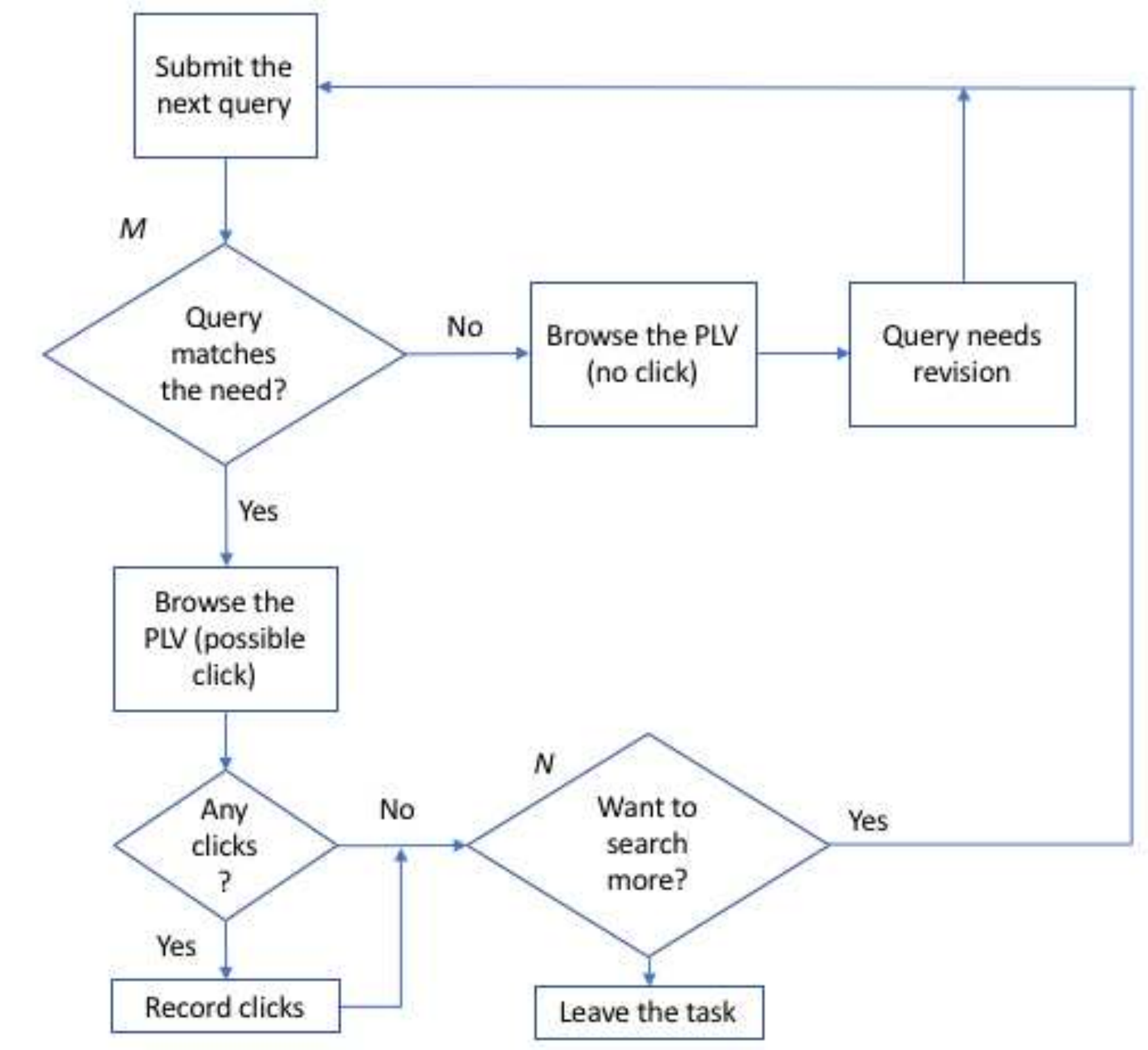}
\caption{Task-centric Click Model with Macro Bernoulli variables $M$, $N$ labelled.  \label{fig:tcm_flow_chart}}
\end{figure}

Instead of using a click model that considers each query-PLV independently,
we will use a form of behavioral analysis that groups search requests into tasks.
The point of view we are taking is similar to the one adopted in \S4 of \cite{zhang2011user}
and subsequent works on the Task-centric Click Model (TCM).  Similar
to the TCM of \cite{zhang2011user}, we assume that when a user searches
on several semantically related queries in the same session, the user goes
through a process of successively querying, examining results,
possibly with clicks, and refining the query until it matches her intent.
We sum up the process in the flow chart, Figure \ref{fig:tcm_flow_chart}, 
which corresponds to both Figure 2, the "Macro Model" and Figure 3, the "Micro model"
in \cite{zhang2011user}. 

\begin{figure}
	\includegraphics[height=2in, width=3in]{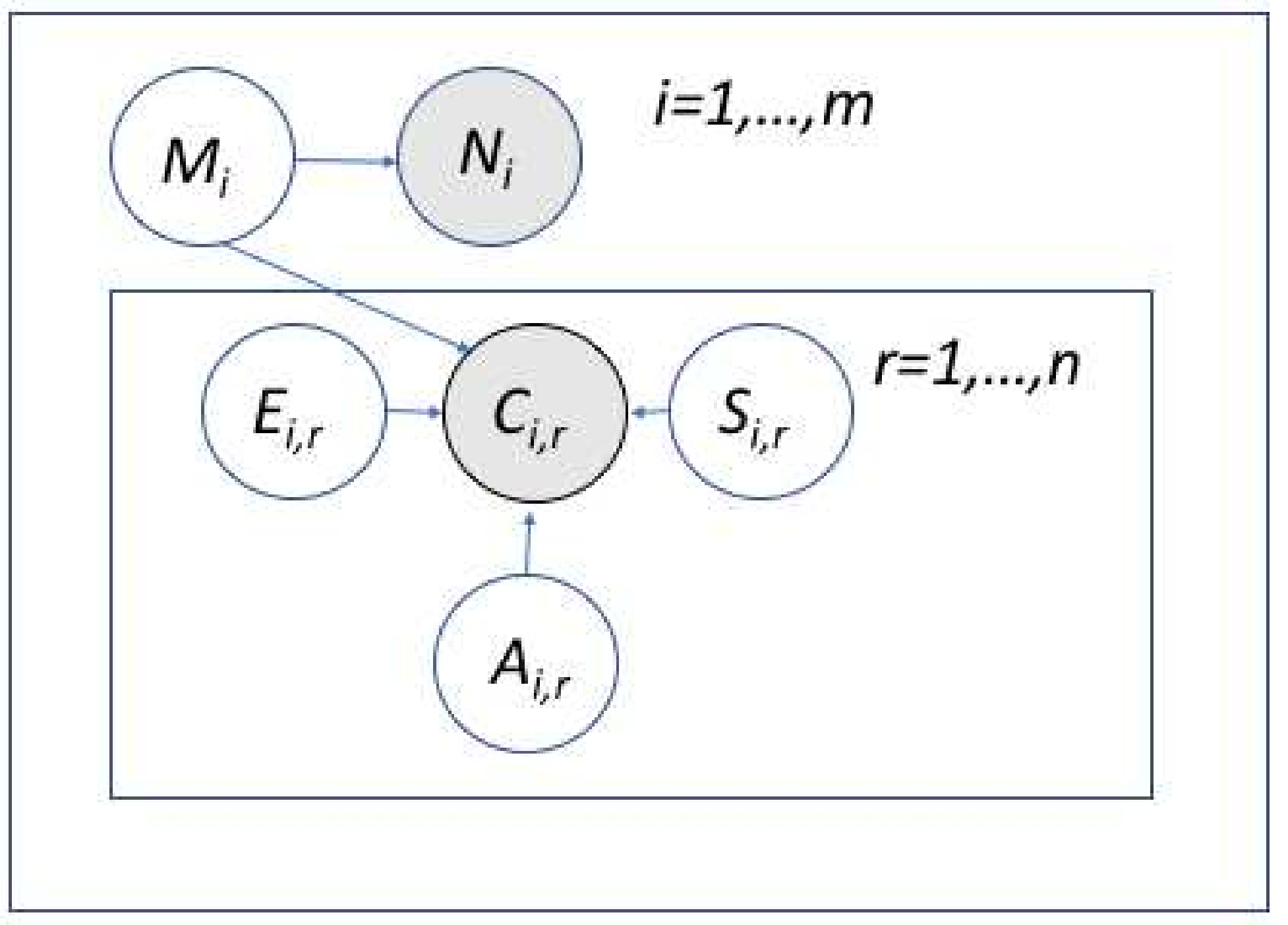}
	\caption{Template-based DAG representation of the TCM \label{fig:tcm_plate_model}}
	\end{figure}

There are several important ways in which we have
simplified the analysis and model as compared with the TCM of \cite{zhang2011user}.
First, we do not consider the so-called "duplicate bias" or the associated freshness
variable of a SKU in our analysis, but we do indicate explicitly that the click
variable depends on both relevance and other factors of SKU attractiveness.
Second, we do not consider the \textit{last}
query in the query chain, or the clicks on the last PLV, as being in any way special.
Third, \cite{zhang2011user} perform inference and parameter tuning on the model,
whereas in this work, at least, we use the model for a different purpose (see below).  As a result \cite{zhang2011user} need to adopt
a particular click model inside the TCM (called the "micro model", or PLV/SERP interaction model) to fully specify
the TCM.  For our purposes the PLV interaction model could remain unspecified, but for the sake of concreteness,
we specify a particular model, similar to the PBM, to govern the user's interaction
with the PLV, inside TCM in Figure \ref{fig:tcm_plate_model},
which is comparable to Figure 4 of \cite{zhang2011user}.  Note that in comparison
to their TCM model we have added the $A$ (attractiveness factor), eliminated the
"freshness" and "previous examination factors", and otherwise just made some 
changes of notation, namely using $S$ instead of $R$ to denote relevance/satisfaction in agreement
with the notation of \S\ref{subsec:click_models}, and the more standard $r$ instead of $j$ for "rank".
The important new variables present in the TCM are the following two relating to the 
\textit{session flow}, rather than the internal search request flow (continuing the numbering (1)--(3) from
\S\ref{subsec:click_models}):
\begin{enumerate}
		\setcounter{enumi}{3}
	\item The user's intent being matched by query $i$, denoted by $M_i$;
	\item  The user submitting another search request after the $i$th query session, denoted by $N_i$.
\end{enumerate}
A complete specification of our TCM consists of the template-based DAG representation 
of figure \ref{fig:tcm_plate_model} together with the following parameterizations (compare (16)--(24) of \cite{zhang2011user}):
\[
\begin{aligned}[t]
P(M_i=1) &= \alpha_1\in[0,1] \\
P(N_i=1|M_i=1) &= \alpha_2\in[0,1]\\
P(E_{i,r})&=\gamma_r\\
P(A_{i,r}) &= \alpha_{u,q}\\
P(S_{i,r})&=\sigma_{u,q}\\
M_i=0&\Rightarrow N_i=1\\
C_{i,r}=1&\Leftrightarrow M_i=1,\, E_{i,r}=1,\, S_{i,r}=1,\, A_{i,r}=1.
\end{aligned}
\]
Resuming our discussion from \S\ref{sec:NIR_models}, we are seeking a way to extract programmatically
from the click logs a collection of triples $(q,d_{\rm rel}, d_{\rm irrel})\in Q\times D\times D$ where ${\sigma_{d_{\rm rel},q}> \sigma_{d_{\rm irrel},q}}$,
which is sufficiently "adversarial".  The notion of "adversarial" which we adopt, is that, first,
$q$ belongs to a \textit{task} (multi-query session) in the sense of the TCM, $q=q_{i'}$ was preceded
by a "similar" query $q_i$, and $d_{\rm irrel}$, while irrelevant to $q_{i'}$, is relevant to $q_{i}$.  Note that
this method of identifying the triples, at least heuristically, has a much higher chance of producing adversarial
example because the similarity
between $q_i$ and $q_{i'}$ implies with high probability that  ${\sigma_{d_{\rm rel},q_{i'}} - \sigma_{d_{\rm irrel},q_{i'}}}$
is much smaller than would be expected if we chose $d_{\rm irrel}$ from the SKU collection \textit{at random}.
Leaving aside the question, for the moment, of how to define search sessions (a point we will return to in Section \ref{sec:dataset_construction}),
we can begin our approach to the construction of the triples by defining criteria which make it likely
that the user's true search intent is expressed by $q_{i'}$, but not by $q_{i}$ (recall again that $i<i'$, meaning $q_i$ is the earlier of the two queries):
\begin{enumerate}
	\item The PLV for $q_i$ had no clicks: $C_{q_i,u_r}=0,\; r=1,\ldots n$.
	\item The PLV for $q_{i'}$ had (at least) one click, on $r'$: $C_{q_{i'},u_{r'}}=1$.
\end{enumerate}
It turns out that relying on these criteria alone is too na\"{i}ve.  The problem is not the relevance $u_{q_{i'},r'}$ to $q_{i'}$,
but the supposed \textit{irrelevance} of the $u_{q_i,r}$ to $q_i$.
In the terms of the TCM, this is because criterion (1), absence of any click on the PLV
for $q_i$, does not in general imply that $M_i=0$.  In \cite{zhang2011user}, the authors address this issue by
performing parameter estimation using the click logs holistically.  We make adopt a version this approach
in future work.  In the present work, we take a different approach, which is to examine the \textit{content}
of $q_{i}$, $q_{i'}$ and add a filter (condition) that makes it much more likely that $M_i=0$, namely
\begin{enumerate}
		\setcounter{enumi}{2}
	\item The tokens of $q_{i'}$ properly contain the tokens of $q_i$.
\end{enumerate}
An example of a $(q_{i}, q_{i'})$ which satisfies (3) is (bookshelf, bookshelf with doors),
whereas an example which does \textit{not} satisfy (3) is (wooden bookshelf, bookshelf with doors):
see the last line of Table \ref{tab:examples}.  The idea
behind (3) is in this situation, the user is \textit{refining} her query to better match
her true search intent, so we apply the term "refinement" either to $q_{i'}$
or to the pair $(q_{i}, q_{i'})$ as a whole.  This addresses the issue that we need $M_i=0$
to conclude that $u_{i,r}$ are likely irrelevant to the query $q_{i'}$.

However, there are still another couple of issues with using the triple  $(q', u_{q_{i'},r'}, u_{q_i,r})$
as a training example.  The first stems from the observation that lack
of click on $u_{q_i,r}$ provides evidence for the irrelevance (to the user's true intent $q'$) 
\textit{only if} it was examined, $E_{u_{q_i,r}}=1$.  This is the same observation as \eqref{eqn:likelihood_implication},
but in the context of TCMs.  We address this by adding a filter:
\begin{enumerate}
	\setcounter{enumi}{3}
	\item The rank $r$ of $u_{q_i,r}\leq \rho$; $\rho$ a small integer parameter,
\end{enumerate}
implying that $\gamma_r$ is relatively close to $1$.  The second is an "exceptional" type of situation
where $M_i=0$, but certain $u_{i,r}$ on the PLV for $q_i$ are relevant to $q_{i'}$.  Consider as an example,
the user issues the query $q_i$ "rubber band", then the query "Acme rubber band" $q_{i'}$, then
clicks on an SKU $u_{q_{i'},r'}=u_{q_i,r}$ she has previously examined on the PLV for $q_i$.
This may indicate that the PLV for $q_i$ actually contained some results relevant to $q_{i'}$, and examining
such results reminded the user of her true search intent. In order to filter out the noise that
from the dataset which would result from such cases, we add this condition:
\begin{enumerate}
	\setcounter{enumi}{4}
	\item $u_{q_{i'},r'}$ does not appear anywhere in the PLV for $q_i$.
\end{enumerate}

\label{sec:task_models}
\section{Dataset Construction}
\begin{table*}
	\begin{tabular}{c m{7cm} m{7cm}}
		\toprule
		Query &                                      Relevant SKU &                                     Irrelevant SKU \\
		\midrule
		epson ink cartridges & Epson 252XL High-capacity Black Ink Cartridge & Canon CL-241XL Color Ink Cartridge \\
		\hline
		batteries aa &  Sony S-am3b24a Stamina Plus Alkaline Batteries (aa; 24 Pk) &  Durcell Quantum Alkaline Batteries --- AAA, 12 Count \\
		\hline
		microsd 128gb &  Sandisk Sdsqxvf-128g-an6ma Extreme Microsd Uhs-i Card With Adapter (128gb) & PNY 32GB MicroSDHC Card \\
		\hline
		accent chair & Acme Furniture Ollano Accent Chair --- Fish Pattern --- Dark Blue & ProLounger Wall Hugger Microfiber Recliner \\
		\hline
		bar stool red 2 & Belleze\copyright\,  Leather Hydraulic Lift Adjustable Counter Bar Stool Dining Chair Red --- Pack of 2 & Flash Furniture Modern Vinyl 23.25 In. --- 32 In. Adjustable Swivel Barstool \\
		\hline
		bookshelf with doors & Better Homes and Gardens Crossmill Bookcase with Doors, Multiple Finishes & Way Basics Eco-Friendly 4 Cubby Bookcase \\
		\bottomrule
	\end{tabular}
	\caption{ Sample of Dataset \label{tab:examples}, showing only original titles for SKU text}
\end{table*}
\label{sec:dataset_construction}
We group consecutive search requests by the same user into one session. Within each session, we extract semantically related
query pairs $(q_i, q_{i'})$, $q_i$ submitted before $q_{i'}$. 
The query $q_{i'}$ is considered semantically related to the previously submitted $q_i$ if they satisfy condition (3) above in 
\S\ref{subsec:task_models}. We keep only $(q_i, q_{i'})$ satisfying (1)--(3) in \S\ref{subsec:task_models}.
As explained in \S\ref{subsec:task_models}, if in addition the clicked SKU $u_{q_{i'}, r'}$,
and the unclicked SKUs $u_{q_i,r}$ satisfy (4)--(5), we have high confidence that $M_i=0$ and $u_{q_i,r}$ does \textit{not} satisfy the user's information need, whereas $M_{i'}=1$ and $u_{q_{i'}, r'}$ does. Therefore, based on our heuristic, we can construct an example for our dataset by  constructing as many as $\rho$ training examples for each eligible click, of the form
\[
(q,d_{\rm rel}, d_{\rm irrel}) := (q_{i'}, \;u_{q_{i'},r'}, \;u_{q_i,r}),\; r=1,\ldots, \rho.
\]

 For our experiments, we processed logged search data on Jet.com from April to November 2017. We filtered to only search requests related to electronics and furniture categories, so as to enable fast experimentation. 
 We implemented our construction method in Apache Spark \cite{zaharia2016apache}.
 Our final dataset consists of around 3.6 million examples, with 130k unique $q$, 131k unique 
$d_{\rm rel}$,  275k unique $d_{\rm irrel}$, with 68k SKUs appearing
as $d_{\rm rel}$ and $d_{\rm irrel}$ for different examples. 
Table \ref{tab:examples} shows some examples extracted by our method.

In order to compensate for the relatively small number of unique $q$ in the dataset and the existence
of concept drift in eCommerce,
we formed the train-validate-split in the following manner rather than the usual
practice in ML of using random splits: we reserved the first six months of data for training,
the seventh month for validation, and the eighth (final) month for testing.  Further, we filtered
out from the validation set all examples with a $q$ seen in training; and from the test
set, all examples with a $q$ seen in validation \textit{or} training.  This turned out to result
in a (training:validation:test) ratio of $(75:2.5:1)$.  Although the split is lopsided towards training
examples, it still results in 46k test examples.  We believe this drastic deletion of validation/test
examples is well worth it to detect overfitting and distinguish true model learning from mere memorization.

We now give an overview of the principles used to form the SKU (document) text as actually encoded by the models.
The query text is just whatever the user types into the search field.  The SKU text is formed
from the concatenating the SKU title with the text extracted from some other fields associated
with the SKU in a database, some of which are free-text description, and others of which are more structured in nature.  Finally, both the query and SKU texts are lowercased and normalized to
clean up certain extraneous elements such as html tags and non-ASCII characters and 
to expand some abbreviations commonly found the SKU text.  For example,
an apostrophe immediately following numbers was turned into the token "feet".  No models
and no stemming or NLP analyzers, only regular expressions, are used in the text normalization.
\section{Experimental Results}
\subsection{Details of Model Training}
All of the supervised IR models were trained on one NVIDIA Tesla K80 GPU using 
PyTorch \cite{paszke2017automatic}, with the \textit{margin ranking loss}:
\[
 L(f(q,d_{\rm rel}), f_(q,d_{\rm irrel})) = \max(0,  f(q,d_{\rm irrel})- f(q,d_{\rm rel}) + 1).
\] 
 We used the Adam optimizer \cite{kingma2014adam},
with an initial learning rate of $1\times 10^{-4}$, using PyTorch's built-in learning rate scheduler
to decrease the learning rate in response to a plateau in validation loss.  For the kernel-pooling
model, it takes about $8$ epochs, with a run time of about $2.5$ hours assuming a batch-size of $512$,
for the learning rate to reach $1\times 10^{-6}$, after which further decrease in the learning
rate does not result in significant validation accuracy improvements.  Also, for the kernel-pooling
model, we explored the effect of truncating the SKU text at various lengths.  In particular,
we tried keeping the first $32$, $64$ and $128$ tokens of the text, and we report the results below.
For the CLSM and Siamese models we tried changing the dimension of both $V$, the distributed
query/document representations, and the number of \textit{channels} of the $\mathrm{cnn}$
layer (dimension of input to $\mathrm{mlp}_1$) using values evenly spaced in logarithmic space
between $32$ and $512$.  We also tried $3$ different values of the dropout rate in these models.

For the Kernel-pooling model, we used word vectors from an unsupervised pre-training step,
using the training/validation texts as the corpus.  As our word2vec algorithm, we used the
CBOW and Skipgram algorithms as implemented
Gensim's \cite{rehurek_lrec} FastText wrapper.  We observed no improvement in performance of the relevance model from changing the choice of word2vec algorithm or altering the word2vec hyperparameters from their default values.
For the tf---idf baseline, we also used the Gensim library, without changing
any of the default settings, to compile the idf (inverse document frequency) statistics. 
 
 \subsection{Error Rate Comparison}
 \begin{table}\centering
 	\begin{tabular}{@{}lll@{}}
 		Model                                                       & Validation & Test\\
 		\toprule
 		Kernel-pooling, trainable embeddings\\
 	    \quad truncation length 32 & 68.63 & 65.62\\
 	    	\quad truncation length 64 & \textbf{63.13} & \textbf{62.92}\\
 		\quad truncation length 128 & 70.16 & 73.97\\
 		\midrule
 		Kernel-pooling, frozen embeddings\\
 		\quad truncation length 64 & 66.13 & 65.40\\
 		\midrule
 		tf---idf baseline    & N/A & 100.0 \\
 		 \bottomrule                    
 		\end{tabular}
 	\caption{Error rates of models, reported as percent of error rate of tf---idf baseline.
 	Validation column reports error rate for the best (lowest) training epoch.  Note
 	that, in all experiments to date on our dataset, none of the distributed
 	models (DSSM, CLSM, Siamese) have outperformed the baseline. \label{tab:error_rates}}
 \end{table}
 We have reported our main results, the error rates of the trained relevance models in Table \ref{tab:error_rates}.
 \textit{The most notable finding is that the kernel-pooling model, our main representative of the 
 distributed class showed a sizable improvement over the baseline, and in our experiments thus far,
 none of the distributed representation models even matched the baseline.}  We found that the 
distributed models had adequate capacity to overfit the training data, but they are not generalizing
well to the validation/test data.  We will discuss ongoing efforts to correct this in \S\ref{sec:conclusions} below.
Another notable finding is that there is an ideal truncation length of the SKU texts for the kernel-pooling
model, in our case around $2^6$ tokens, which allows the model enough information without introducing too much extraneous noise.  
Finally, following \cite{xiong2017end}, we also evaluated a variant of the kernel-pooling model
where the word embeddings were "frozen", i.e. fixed at their initial word2vec values.  Interestingly,
unlike what was observed in \cite{xiong2017end}, we observed only a modest degredation
in performance, as measured by overall error rate, from the model using the word2vec embeddings as compared with the full model.
Based on the qualitative analysis of the learned embeddings, \S\ref{sec:fine_tuned_embeddings},
we believe it is still worthwhile to train the full model. 
	\begin{table*}
		\centering
		\begin{tabular}{@{}cc@{}}
			\includegraphics[width=.45\textwidth]{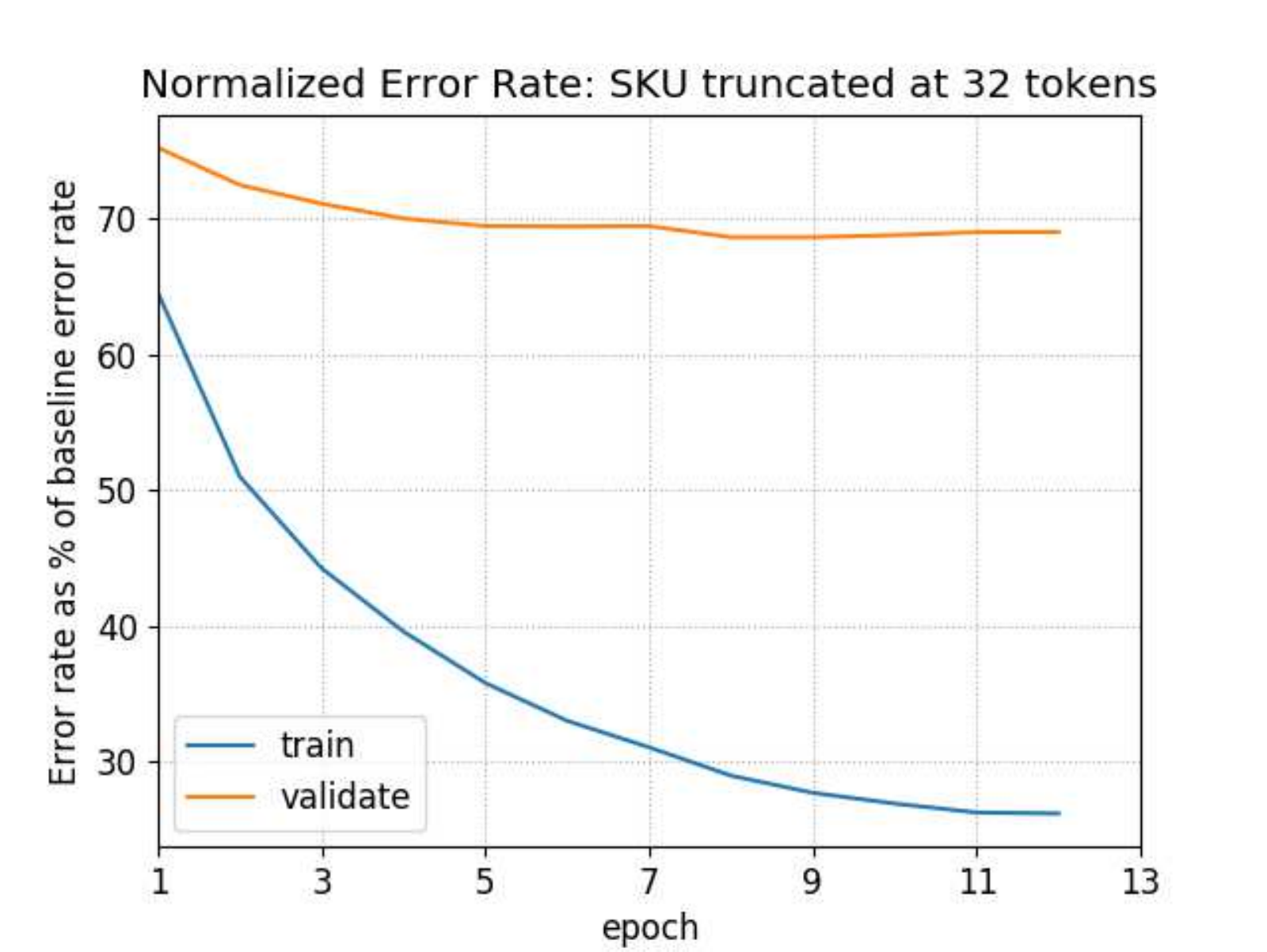} &
			\includegraphics[width=.45\textwidth]{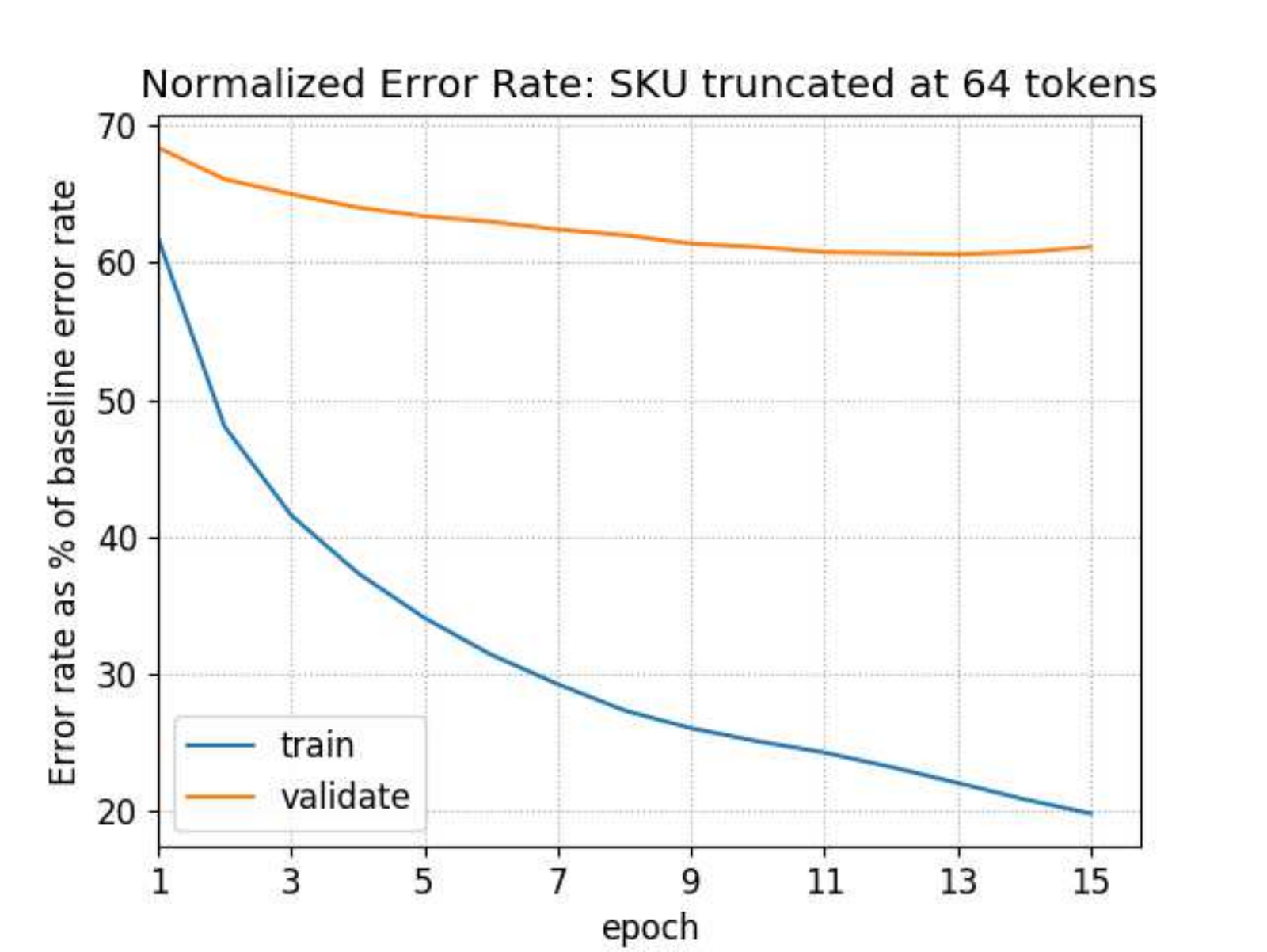} \\
				\includegraphics[width=.45\textwidth]{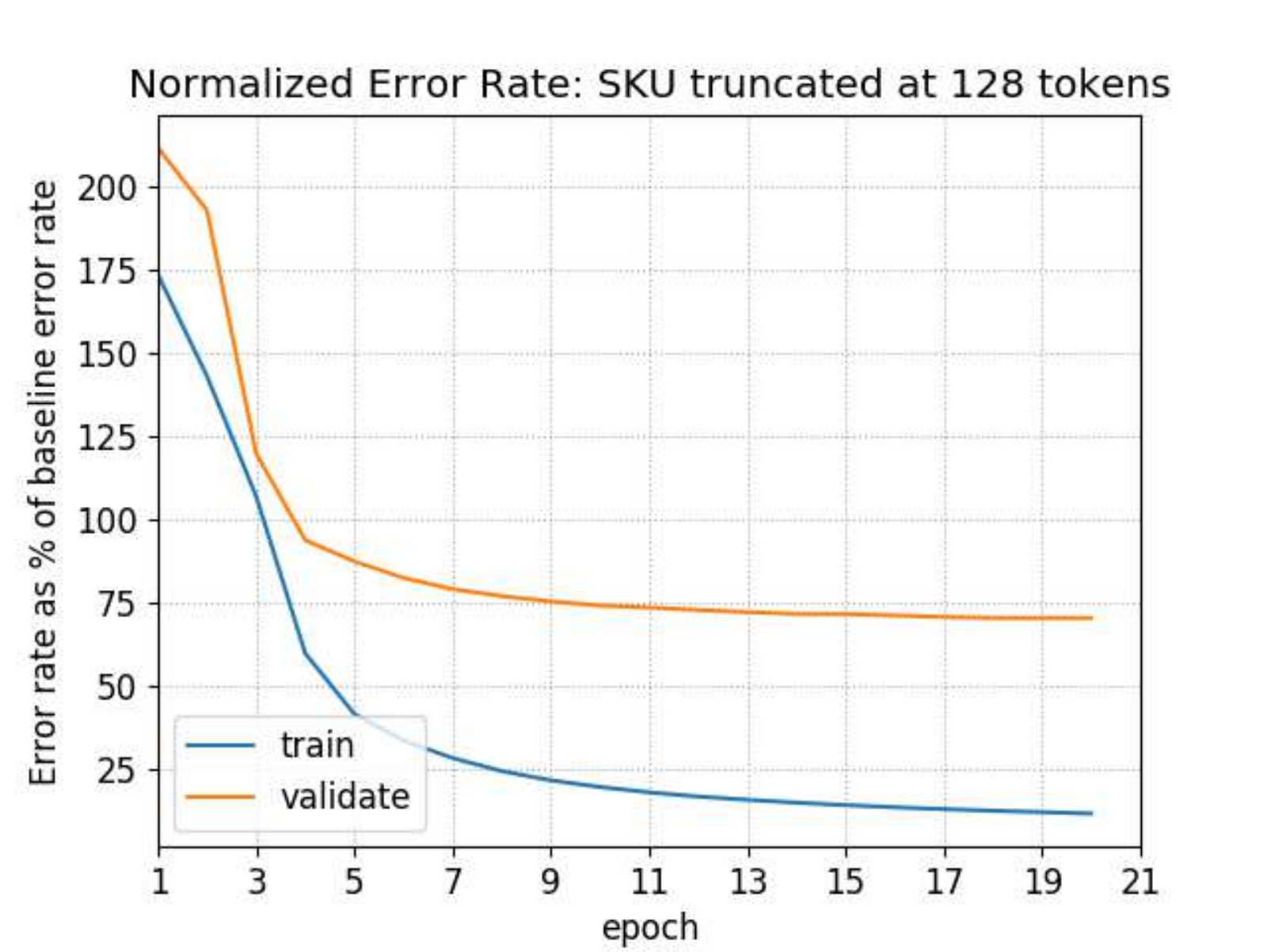} &
			\includegraphics[width=.45\textwidth]{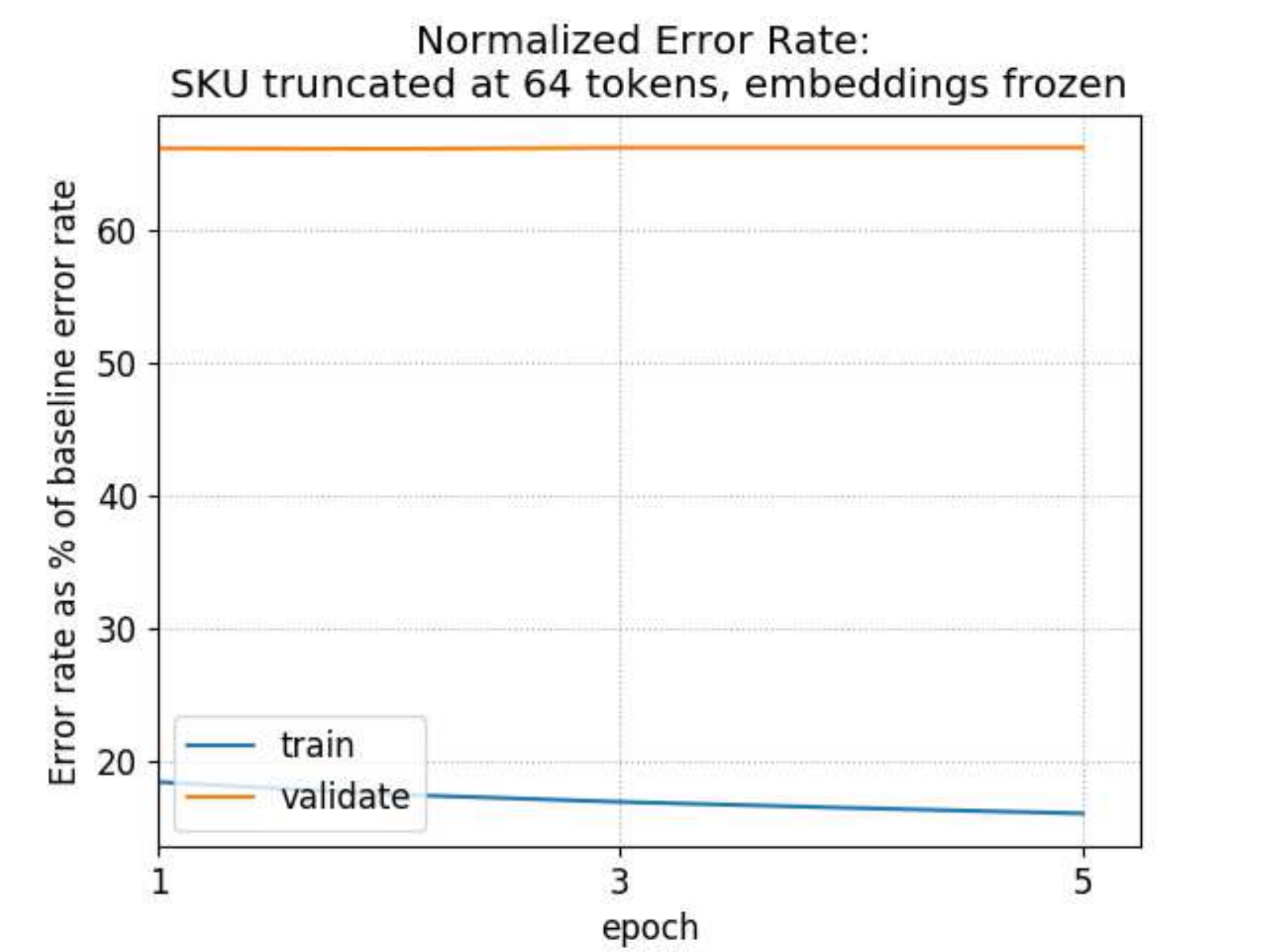}   \\
		\end{tabular}
		\caption{Training paths of Kernel Pooling Model With Different Hyperparameter Choices \label{sec:experimental_results}}
	\end{table*}
\subsection{Pre-trained versus fine-tuned embeddings}
\label{sec:fine_tuned_embeddings}
Similarly to \cite{xiong2017end}, we found that the main effect of the supervised retraining of the 
word embeddings was to decouple certain word pairs.  Corresponding to Table 8 of their
paper we have listed some examples of the moved word pairs in Table \ref{tab:moved_word_pairs}.
The training decouples roughly twice as many word pairs as it moves closer together.  In spite
of the relatively modest gains to overall accuracy from the fine-tuning of embeddings, we believe
this demonstrates the potential value of the fine-tuned embeddings for other search-related tasks.
\begin{table}\centering
	\begin{tabular}{@{}lll@{}}
From & To & Word Pairs\\	
\toprule
$\mu=0.8$ & $\mu=0.1$ & (male, female), (magenta, cyan) \\ 
$\mu=0.5$ & $\mu=0.1$ & (64gb, 128gb), (loveseat, pillows) \\ 
$\mu=0.1$ & $\mu=-0.3$ & (piece, sensations), (monitor, styling) \\ 
$\mu=0.5$ & $\mu=-0.1$ & (internal, external), (tv, hdtv) \\ 
$\mu=-0.1$ & $\mu=0.3$ & (vintage, component), (replacement, pedestal) \\ 
\bottomrule
	\end{tabular}
\caption{Examples of moved word pairs \label{tab:moved_word_pairs}}
\end{table}
\section{Conclusions}
We showed how to construct a rich, adversarial dataset for eCommerce relevance.  We demonstrated
that one of the current state-of-art NIR models, namely the Kernel Pooling model, is able to reduce
pairwise ranking errors on this dataset, as compared to the tf---idf baseline, by over a third.  We
observed that the \textit{distributional} NIR models such as DSSM and CLSM overfit and do 
not learn to generalize well on this dataset.  Because of the inherent advantages of distributional
over local-interaction models, our first priority for ongoing work is to diagnose and overcome this overfitting
so that the distributional models at least outperform the baseline.  The work is proceeding along
two parallel tracks.  One is explore further architectures in the space of all possible NIR models to find
ones which are easier to \textit{regularize}.  The other is to perform various forms of \textit{data augmentation},
both in order to increase the sheer quantity of data available for the models to train on and to overcome
any biases that the current data generation process may introduce.
\label{sec:conclusions}
\begin{acks}
The authors would like to thank Ke Shen for his assistance
setting up the data collection pipelines.

%
%
\end{acks}

\bibliographystyle{ACM-Reference-Format}
\bibliography{sample-bibliography}

\end{document}